\documentclass[
  aps,
  prx,
  reprint,
  twocolumn,
  amsmath,amssymb,
  superscriptaddress,
  longbibliography
]{revtex4-2}

\usepackage{graphicx} 
\usepackage{float}
\usepackage{amsmath}
\usepackage{amssymb}
\usepackage{amsfonts}
\usepackage{euscript}
\usepackage{enumerate}
\usepackage{hhline}
\usepackage{pslatex}
\usepackage{tabularx}
\usepackage{braket}
\usepackage{soul}
\usepackage[dvipsnames]{xcolor}

\newcommand{\ysnote}[1]{{\color{red} #1}}

\makeatletter
\renewcommand{\@dotsep}{500}
\renewcommand{\@pnumwidth}{0em}
\renewcommand{\l@figure}[2]{
\newcommand{\ysnote}[1]{{\color{red} #1}}
\@dottedtocline{1}{1.5em}{2em}{Figure #1}{}\vspace{15pt}}
\usepackage[colorlinks,linkcolor=red,anchorcolor=blue,citecolor=blue]{hyperref}
\begin{document}

\title{Engineering Biquadratic Interactions in Spin-1 Chains by Spin-1/2 Spacers}

\author{Yasser Saleem}
\thanks{These authors contributed equally to this work.}
\affiliation{Condensed Matter Theory, Department of Physics, TU Dortmund, 44221 Dortmund, Germany.}
\affiliation{
Institute for Theoretical Physics and Astrophysics,
and W\"urzburg-Dresden Cluster of Excellence on Complexity and Topology in Quantum Matter ct.qmat,
Julius-Maximilians-Universität W\"urzburg, Am Hubland, D-97074 W\"urzburg, Germany}

\author{Weronika Pasek}
\thanks{These authors contributed equally to this work.}
\affiliation{Institute of Physics, Faculty of Physics, Astronomy and Informatics, Nicolaus Copernicus University, Grudziadzka 5, 87-100 Toru\'n, Poland}

\author{Marek Korkusinski}
\affiliation{Department of Physics, University of Ottawa,
Ottawa K1N6N5, Canada}
\affiliation{Security and Disruptive Technologies,
National Research Council, Ottawa K1A0R6, Canada }
\author{Moritz Cygorek}
\affiliation{Condensed Matter Theory, Department of Physics, TU Dortmund, 44221 Dortmund, Germany.}
\author{Pawe\l ~Potasz}
\affiliation{Institute of Physics, Faculty of Physics, Astronomy and Informatics, Nicolaus Copernicus University, Grudziadzka 5, 87-100 Toru\'n, Poland}

\date{\today}

\begin{abstract}
\noindent Low-dimensional quantum systems host a variety of exotic states, such as symmetry-protected topological ground states in spin-1 Haldane chains. Real-world realizations of such states could serve as practical quantum simulators for quantum phases if the interactions can be controlled. However, many proposed models, such as the Affleck-Kennedy-Lieb-Tasaki (AKLT) state, require unconventional forms of spin interactions beyond standard Heisenberg terms, which do not naturally emerge from microscopic (Coulomb) interactions. Here, we demonstrate a general strategy to induce a biquadratic term between two spin-1 sites and to tune its strength $\beta$ by placing pairs of spin-1/2 spacers in between them. $\beta$ is controlled by the ratio of the Heisenberg couplings between the spin-1 sites and the spacer spins, and between the spacer spins themselves. Increasing this ratio increases the magnitude of $|\beta|$ and decreases the correlation length of edge states. Detailed atomistic calculations reveal that chains of nanographene flakes with 22 and 13 atoms, respectively, which could be realized by state-of-the-art bottom-up growth technology, yield precisely the couplings required to approach the AKLT state. These findings deliver a blueprint for engineering unconventional interactions in bottom-up synthesized quantum simulators.
\end{abstract}

\maketitle
 \begin{figure*}[t]
    \centering    \includegraphics[width=0.98\linewidth]{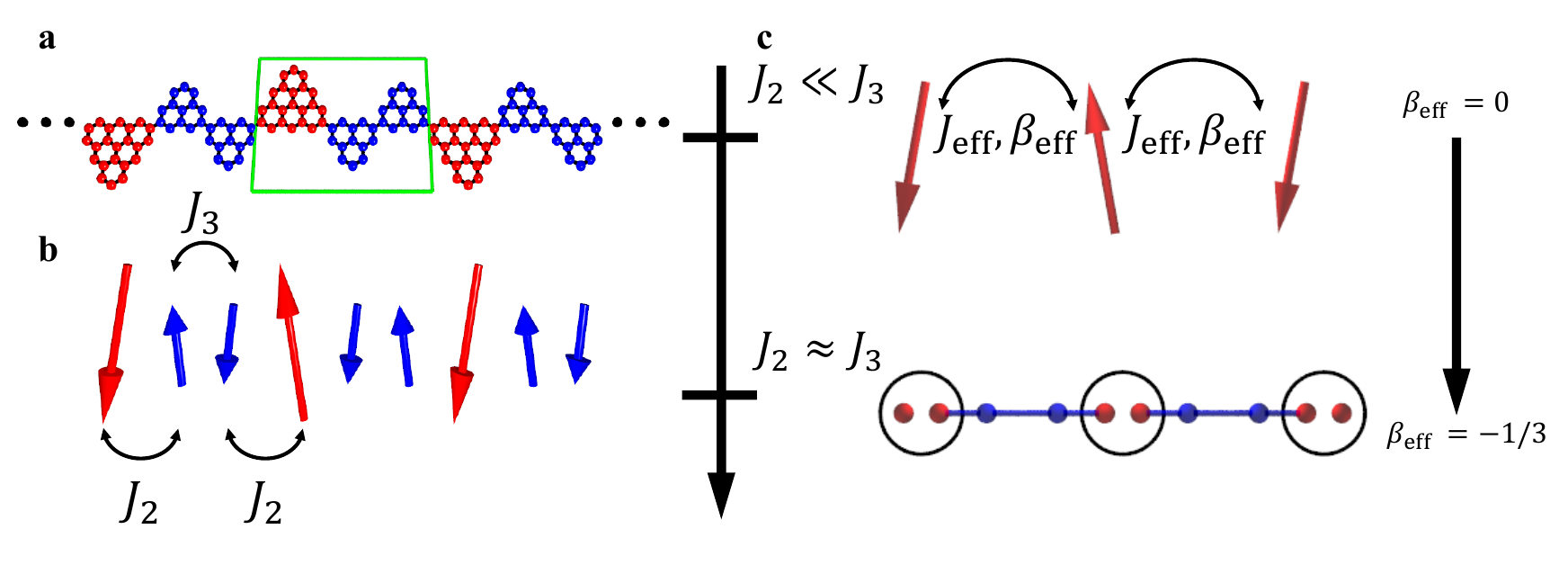}
    \caption{(a) Atomistic nanographene chain, where larger red triangles host localized spin-1 moments and smaller blue triangles host spin-$\tfrac{1}{2}$ moments. 
    (b) Effective spin model with antiferromagnetic exchange couplings $J_{2}$ and $J_{3}$. 
    (c) Ground-state representations of the effective spin-1 chain for $J_{2} \ll J_{3}$ and $J_{2} \approx J_{3}$. 
    Red and blue balls represent effective spin-$\tfrac{1}{2}$ objects; 
    black circles denote ferromagnetic coupling into spin-1 units; 
    solid blue lines indicate singlet bonds; 
    and dotted blue lines represent mixed singlet-triplet bonds.}
    \label{fig:fig1}
\end{figure*}
\noindent Spin lattice systems exemplify the wealth of fascinating nonclassical phenomena of quantum many-body physics. These include quantum phase transitions to exotic phases, such as symmetry-protected topological phases with the emergence of spin fractionalization and quasiparticle formation at the edges of spin chains \cite{ChenPRB2011,TasakiKennedy1992PRB,pollmann2010entanglement,tasakibook}. In particular, such phases lack a local order parameter and thus defy the Landau paradigm of condensed matter physics. Given the plethora of possible arrangements of spins in lattices, general statements are hard to come by. The Haldane conjecture \cite{HALDANE1983Continuum,Haldane1983prl} is one of the few exceptions as it posits quite generally that spin chains consisting of integer spin sites differ qualitatively from their half-integer spin counterparts in that the ground state is usually separated from the excited states by a finite gap, which persists into the thermodynamic limit. In this sense, energy quantization as well as topology, fractionalization, and hidden long-range antiferromagnetic order prevail in macroscopic systems.

Still, most of these fascinating insights have been theoretical in nature.
Experimental quantum simulators based on ultracold atoms in optical lattices have provided a powerful route to emulate model Hamiltonians such as the Hubbard or Heisenberg models, and have even realized symmetry-protected topological phases in spin chains~\cite{esslinger2010fermi,lewenstein2012ultracold,UltraColdBloch2017,Sompet}.
However, experimental investigations of such phenomena in solid-state systems typically rely on the accidental occurrence of specific spin arrangements in natural crystals \cite{renard2002haldane,williams2020near}.
A major challenge for realizing many of the theoretically proposed systems in practice is that only a limited set of interactions—most commonly antiferromagnetic nearest-neighbor couplings—emerge robustly from underlying microscopic exchange mechanisms.

Here, we demonstrate a strategy to engineer more unconventional forms of spin-spin interactions from microscopic interactions. Our target is inducing and controlling the $\beta$ parameter determining the strength of the bilinear-biquadratic (BLBQ) term in the Hamiltonian of an effective spin-1 chain
\begin{equation}
 H_{BLBQ}= J\sum\limits_{i}\left[\Vec{S}_i\cdot\Vec{S}_{i+1}-\beta\left(\Vec{S}_i\cdot\Vec{S}_{i+1}\right)^2\right],
\label{eq:BLBQ}
\end{equation}
because of the extraordinary importance of the Affleck-Kennedy-Lieb-Tasaki (AKLT) state in quantum many-body theory, which is realized by the ground state for $\beta=-1/3$ \cite{affleck1987,affleck1988valence}. Here, $J$ defines the overall energy scale while $\beta$ represents the ratio between BLBQ and Heisenberg terms. In fact, the BLBQ model has a rich phase diagram with a gapped Haldane phase in the parameter range $-1<\beta<1$, which includes the Heisenberg point $\beta=0$ \cite{Solyom1987,LauchliBetaPhaseDiagramPrb2006}. This gap, as well as the emergence of a quadruply degenerate ground state in open chains that becomes exact in the thermodynamic limit and corresponds to spin-1/2 quasiparticles localized at the edges \cite{kennedy1990exact,White1992,WhiteHuseRGs1PRB1993}, has come to be understood as a manifestation of a symmetry-protected topological quantum phase \cite{ChenPRB2011,pollmann2010entanglement,Schuch2011SPTclassif,Pollmann2012}. The AKLT point is special as its ground state can be constructed explicitly in a matrix product state representation and the correlation length of the edge states $\xi=\ln(3)^{-1}\approx 0.91$ is the shortest within the Haldane phase \cite{kennedy1990exact,fannes1992finitely,SCHOLLWOCK1996,SCHOLLWOCKDMRG2011}. This minimum in the correlation length leads to maximally localized, spin-1/2 (fractionalised) edge states and thus remains well defined even in relatively short chains, making the AKLT state the most robust realization of the Haldane phase and an ideal benchmark for identifying and realizing symmetry-protected topological order in quantum spin systems.

Our first main finding is that a finite effective BLBQ interaction can be induced purely from nearest-neighbor antiferromagnetic interactions by placing pairs of spin-1/2 spacers between two spin-1 sites as depicted in Fig.~\ref{fig:fig1} (b). The BLBQ parameter $\beta$ is controlled by the ratio $r=J_2/J_3$ of the coupling constant $J_2$, between spin-1 and spin-1/2, to the coupling $J_3$, between the spin-1/2 spacer sites. For small ratios $r$, the spacer pairs are locked in a singlet state. The effective BLBQ interaction is understood in this regime as mediated by fourth-order perturbative virtual excitations of spacer triplet states, suggesting large ratios are required to approach the AKLT limit. 

A second key finding is motivated by recent experimental realizations for spin-1 chains and spin-1/2 chains formed by nanographene flakes grown with a precise carbon atom number using bottom up techniques \cite{FaselNaturespinchain,zhao2024tunable,Fu2025,yuan,Su2025spinonehalf,Sun2025OnSurfacespinonehalf}. The total spin of a nanographene flake is related to its number of atoms \cite{Rossier2007,GuccluPRL2009,Potasz2010ZeroEnergy,potasz2012} and is protected by Lieb's theorem \cite{LiebTheorems}. In Ref.~\onlinecite{FaselNaturespinchain}, scanning tunneling microscopy provided experimental evidence of edge states from the Haldane phase in spin-1 chains of nanographene flakes, presumably close to the Heisenberg point $\beta\approx 0$. Using detailed atomistic modeling, here, we predict that 22-atom nanographenes separated by spacers of pairs of 13-atom nanographenes (see Fig.~\ref{fig:fig1}(a)) provide antiferromagnetic interactions of precisely the right ratio to realize the AKLT state — large enough to induce an effective $\beta\approx -1/3$. Thus, nanographene realizations of the AKLT state with high fidelity seem feasible with state-of-the-art technology.

\section{Results}
\subsection{From hybrid spin-1/spin-$\frac 12$ chain to BLBQ model}
\noindent Our goal is to show how an effective BLBQ model arises from an antiferromagnetic hybrid spin chain composed of spin-1 and spin-$\frac 12$ sites as depicted in Fig.~\ref{fig:fig1}(b). The Hamiltonian for the hybrid spin chain with $N_s$ spin-$\frac 12$ spacers between each of the $L$ spin-1 sites is 
\begin{equation}
    H =H_{J_2}+H_{J_3},
       \label{eq: SpinModel}
\end{equation}
with
\begin{subequations} \label{eq:H_pert}
\begin{align}
&H_{J_2}=J_2\sum_{m=1}^{L-1}\big[\textbf{S}_{m}\cdot \textbf{s}_{m,1}+\textbf{s}_{m,N_s}\cdot \textbf{S}_{m+1}\big]
\label{eq:HJ2} \\
&H_{J_3}=J_3\sum_{m=1}^{L-1}\sum_{i=1}^{N_s-1}\textbf{s}_{m,i}\cdot \textbf{s}_{m,i+1}
\label{eq:HJ3} 
\end{align}
\end{subequations}
where $\textbf{S}_{m}$ is the $m$-th spin-1 operator and $\textbf{s}_{m,i}$ is the $i$th spin-$\frac 12$ operator to the right of the $m$th spin-1. We consider finite chains terminated by spin-1 at each end, thus the number of spin-1's is $L$, and $L-1$ is the number of unit cells.
$J_2$ and $J_3$ are coupling constants between spin-1 and spin-$\frac 12$ sites, and between two spin-$\frac 12$ sites, respectively.

Motivated by the fact that antiferromagnetic couplings arise when such spins sites are realized by nanographene flakes [see Fig.~\ref{fig:fig1}(a)], we consider here only the case $J_2,J_3>0$. In our analysis, we consider two distinct cases [see 
Fig.~\ref{fig:fig1} (c)].
In the regime $J_2\ll J_3$, the spacer spins are essentially locked in singlets but mediate an effective interaction between two spin-1's, which can be understood by perturbation theory on a single spin-1 dimer with a pair of spin-$\frac 12$ spacers in between. 
Numerical density matrix renormalization group (DMRG) simulations \cite{White1992} are performed to analyze spin chains for $J_2\ll J_3$ up to $J_2 > J_3$, where we find ground states consistent with an effective BLBQ coupling evolving from the Heisenberg point $\beta_\text{eff}\approx 0$ for $J_2\ll J_3$ to the AKLT value  $\beta_\text{eff}\approx -\frac 13$ for intermediate $J_2\approx J_3$.

\subsection{Perturbation theory in the limit $J_2\ll J_3$}
\begin{figure}[ht]
    \centering
    \includegraphics[width=1\linewidth]{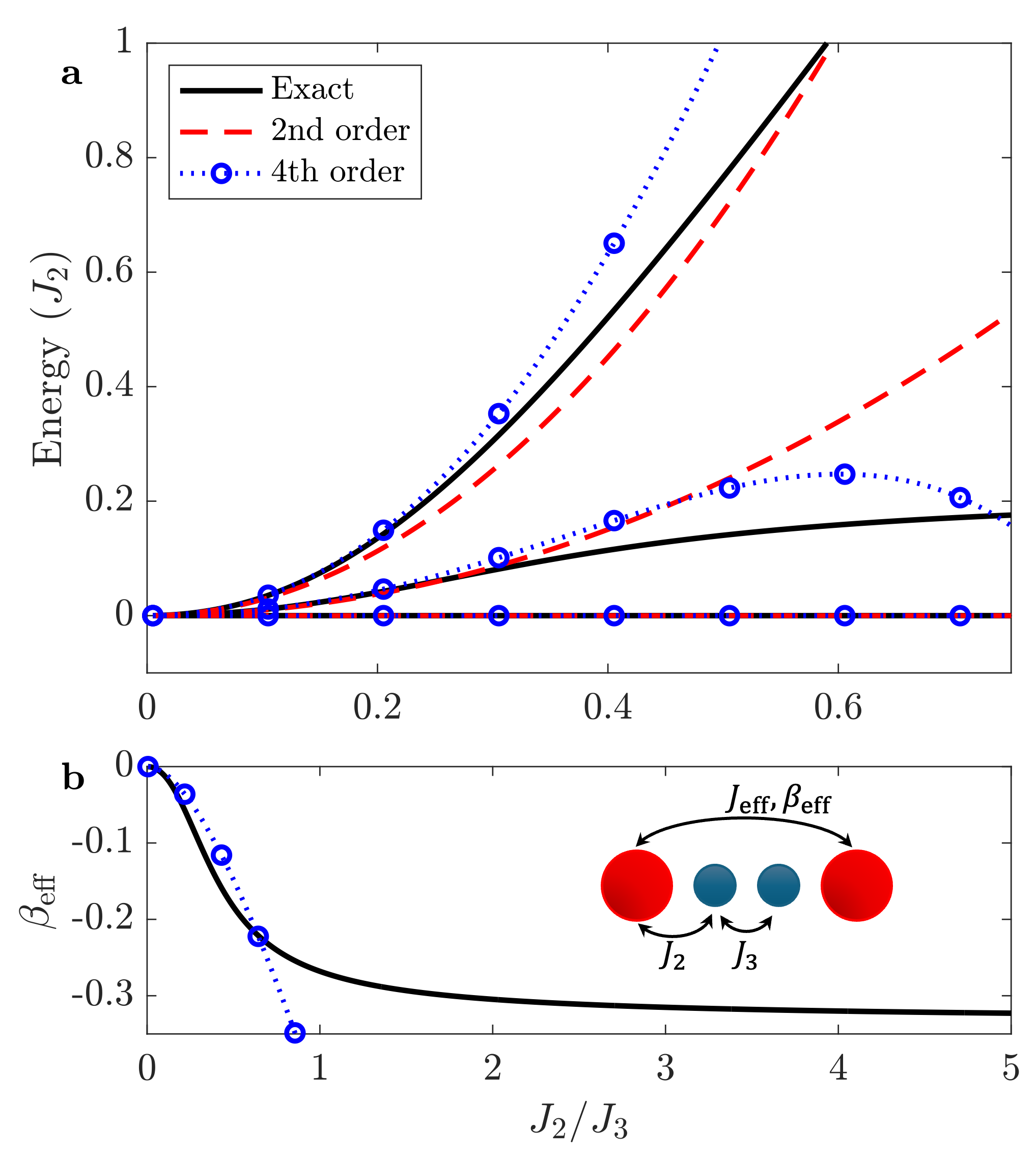}
    \caption{(a) Low-energy spectrum of the hybrid spin-1/spin-$\frac{1}{2}$ chain for a dimer with $N_s=2$, i.e., a spin-1-$\frac 12$-$\frac 12$-1 configuration, as a function of $J_2/J_3$, comparing exact diagonalization results (black) with second-order (red dashed) and fourth-order (blue dotted) perturbation theory. The three lowest eigenstates are shown, corresponding to the effective spin-1 manifold.
    (b), Effective interaction parameter $\beta_{\text{eff}}$ for the hybrid chain, extracted from fitting the spectrum to the two-site bilinear-biquadratic (BLBQ) from Eq.~\ref{eq:BLBQ}, plotted as a function of $J_2/J_3$. The fourth-order perturbative result (blue circles) agrees well with the exact result (black) at small $J_2/J_3$, but deviates as spacer triplet excitations enter the low-energy manifold. }
    \label{fig:fig2}
\end{figure}
The emergence of a spacer-mediated effective BLBQ interaction between two spin-1 sites can be understood most easily in the limit $J_2\ll J_3$ for a single spin-1 dimer, i.e. a spin-1-$\frac 12$-$\frac 12$-1 configuration as depicted in the inset of Fig.~\ref{fig:fig2}(b). When $J_3$ dominates, the spacer spins are locked in a singlet state separated from the triplet states by an energy $J_3$. Eliminating the latter by a Schrieffer-Wolff transformation (see Appendix.~\ref{appsub:MethodsPT}), to fourth order, yields a perfect mapping to an effective BLBQ Hamiltonian in Eq.~\ref{eq:BLBQ} with an effective exchange coupling given by $J_{\mathrm{eff}}\approx\frac{J_2^2}{2J_3}+\frac{3J_2^3}{4J_3^2}-\frac{J_2^4}{8J^3_3}$. A biquadratic term appears only at fourth order with $\beta_{\text{eff}} J_{\text{eff}}=\frac{J_2^4}{2J_3^3}$.  As a side note, for small $J_2/J_3$, with $J_2<0$ (ferromagnetic), an effective BLBQ Hamiltonian still appears with the $\beta_{\text{eff}}$ and $J_{\text{eff}}$ terms having the same signs as in the $J_2>0$ (antiferromagnetic) case.

The range of validity of perturbation theory is visible from Fig.~\ref{fig:fig2}(a), where the low-energy spectrum of this dimer obtained by exact diagonalization is compared to perturbative results to second and fourth order. The three lowest energy states correspond to a singlet, triplet, and quintuplet, analogously to what is expected from solving the BLBQ Hamiltonian in Eq.~\ref{eq:BLBQ} for two sites. Fourth-order perturbative and exact results match well for small values of $J_2/J_3\lesssim 0.3$. Deviations appear first in the quintuplet state for second-order perturbation theory. To extract an effective value for the BLBQ coupling $\beta_\text{eff}$ beyond the validity of perturbation theory, we fit the spectra obtained from exact diagonalization to solutions of the BLBQ Hamiltonian for two sites.
The effective BLBQ coupling is depicted in Fig.~\ref{fig:fig2}(b) as a function of $J_2/J_3$ together with the fourth-order perturbative expression. Results from exact diagonalization suggest that the magnitude of $\beta_\text{eff}$ first increases quadratically from 0 for $J_2/J_3\approx 0$ before $\beta_\text{eff}$ asymptotically approaches the AKLT value of $-\frac 13$ at large $J_2/J_3$.
Perturbation theory only accurately captures the initial increase.
This perturbative analysis of the spin-1 dimer qualitatively demonstrates spacer-mediated spin-spin interactions as a general concept for engineering effective BLBQ interactions, while accurate quantitative values may require numerical simulations. 

 Note that by considering hybrid spin chains instead of just a dimer, further neighbor terms will appear in the perturbation theory analysis. In particular, next-nearest-neighbor magnetic couplings and three-site interaction terms of the form $(\textbf{S}_i\cdot\textbf{S}_{i+1})(\textbf{S}_{i+1}\cdot \textbf{S}_{i+2})$ can appear in fourth order and do not vanish. These terms are analogously derived in the simpler case of a spin-1 chain formed from nanographene dimers in Ref. \onlinecite{Henriques2023t3paper} [see Eq. (9) therein].  As discussed in Appendix C of Ref. \onlinecite{Henriques2023t3paper} , such additional contributions lead to quantitative modifications provided the ratio perturbative coupling does not become too large.
\begin{figure*}
    \centering
    \includegraphics[width=1\linewidth]{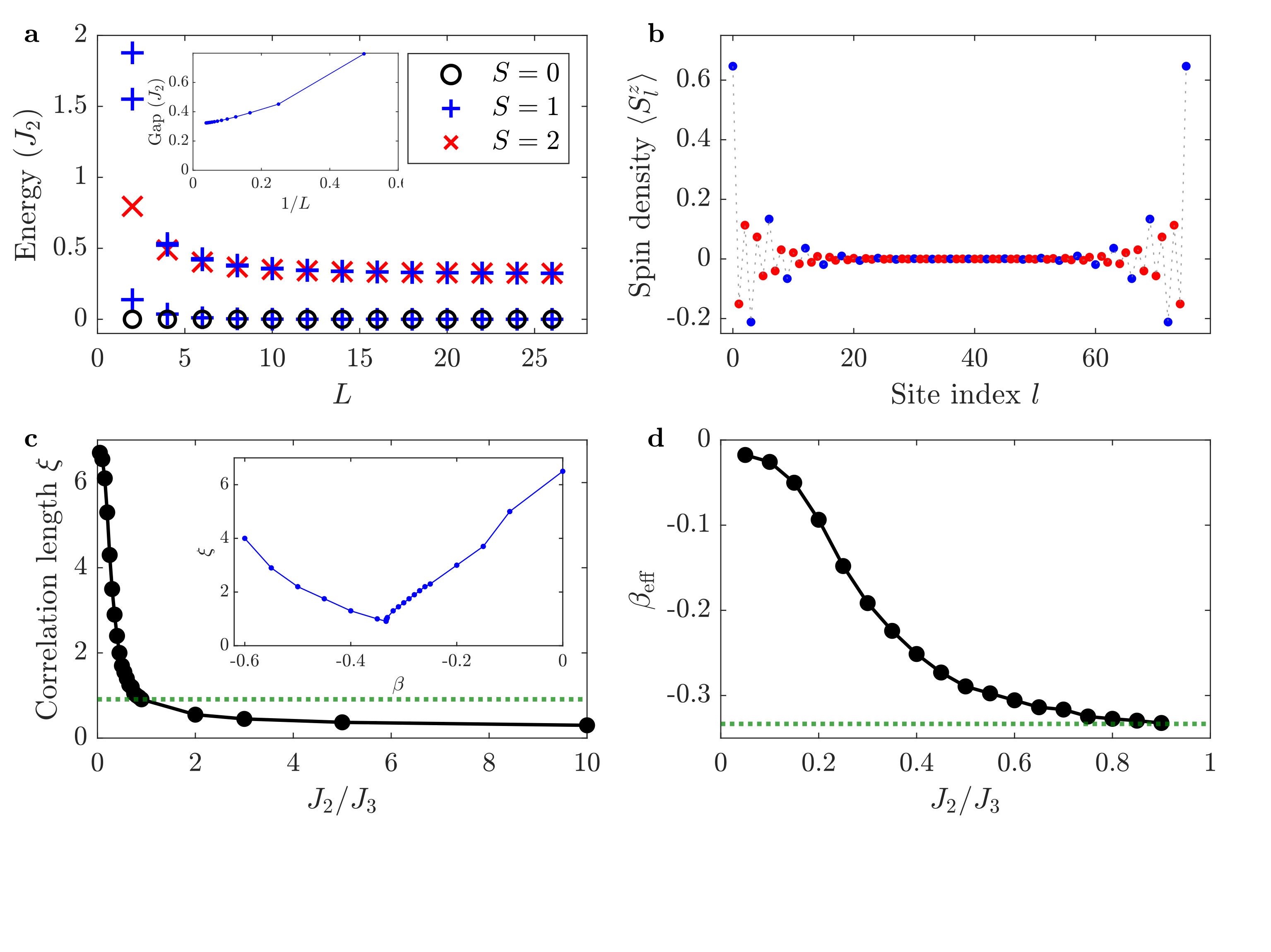}
    \caption{ DMRG results for the hybrid spin-1/spin-$\frac 12$ model with $N_{\rm s}=2$. (a) Low-energy spectrum as a function of the chain length for the Hamiltonian given by Eq. (\ref{eq: SpinModel}) with $J_2/J_3 \sim 0.5$.  The ground state is formed by total spin states $S=0$ and $S=1$, and is degenerate in the thermodynamic limit. The inset shows the Haldane gap converging to a finite value in the thermodynamic limit. (b) The spin density of the triplet (with $S_z=1$) for $L=26$. An index $l=3(m-1)+i$, where $m$ is a unit cell index and $i=0,1,2$, with $i=0$ for a spin-1 and $i=1,2$ for spin-$\frac 12$'s. (c) The correlation length as a function of $J_2/J_3$ for $L=60$. The inset shows the correlation length of the BLBQ Hamiltonian [Eq.~\ref{eq:BLBQ}] as a function of $\beta$. (d) The effective $\beta_{\mathrm{eff}}$ obtained by matching the correlation length of the hybrid spin-1/spin-$\frac 12$ model to that of the BLBQ model.}
    \label{fig:fig3}
\end{figure*}
\begin{figure}
    \centering
    \includegraphics[width=1\linewidth]{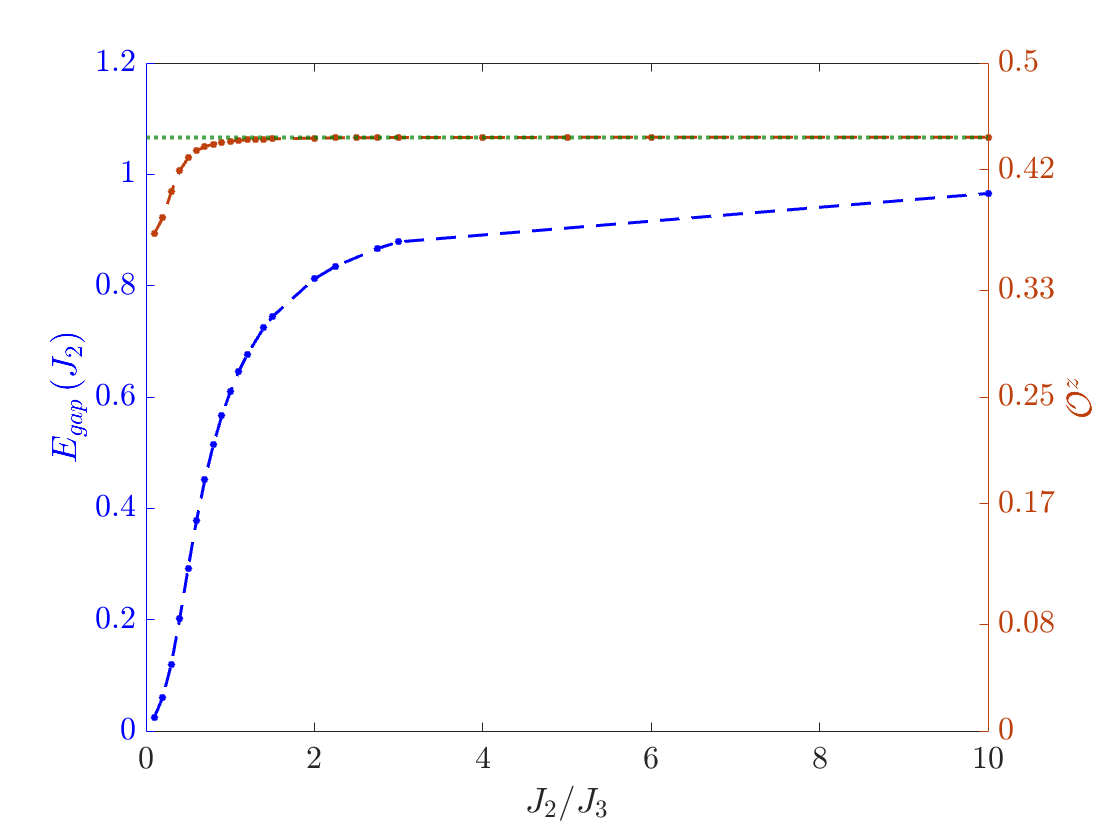}
    \caption{ The evolution of the excitation gap $E_{\mathrm{gap}}$ (blue, left axis) and the SOP $\mathcal{O}^z$ (orange, right axis), as a function of $J_2/J_3$, with the green dotted line denoting the value of the SOP for the AKLT state.}
    \label{fig:fig4}
\end{figure}
\subsection{DMRG analysis of the hybrid spin-1/spin-$\frac 12$ chain}
\noindent Having discussed the principle of spacer-mediated BLBQ interactions on the example of a single spin-1 dimer, we now turn to finite chains. Numerical simulations are performed using DMRG \cite{White1992,SCHOLLWOCKDMRG2011} (see Appendix~\ref{appsub:DMRG}).
We concentrate on the intermediate regime $J_2/J_3=0.5$, where the picture of spacer-mediated BLBQ interaction is qualitatively correct while at the same time the BLBQ coupling is sizable.

In Fig.~\ref{fig:fig3}(a), we show that the low-energy spectrum of the hybrid spin model indeed behaves like that of a spin-1 chain in the Haldane phase: with increasing chain length, singlet and triplet states quickly become degenerate ground states, which is consistent with fractionalised spin-1/2 quasiparticles on either end of the chain. These are separated from excited states by a Haldane gap that remains finite in the thermodynamic limit. 
The spin density of the lowest triplet state with $S_z=1$ for a chain with $L=26$ is shown in Fig.~\ref{fig:fig3}(b).
Even in the range of intermediate ratios $J_2/J_3$, the spin density is predominantly concentrated on the spin-1 sites. Specifically, the values on the first and last sites are above 0.6. Note that this value is $\frac 23\approx 0.66$ for the AKLT state~\cite{polizzi1998}, suggesting that the state has a short correlation length. Indeed, as we show below, the calculated correlation length for $J_2/J_3\sim0.5$ is closer to the one for the AKLT state than for the ground state of a pure Heisenberg antiferromagnetic model.

The ground state of the BLBQ Hamiltonian is within the Haldane phase for $|\beta|<1$, with the remarkably short correlation length in the AKLT state, $\beta=-\frac{1}{3}$. 
That is, the spin operators for the AKLT state $\langle S_{m'}^z S_m^z \rangle=\frac{4}{3}(-1)^{|m-m'|}e^{-|m-m'|\ln{3}}$, gives rise to a correlation length $\xi = \ln(3)^{-1} \approx 0.91$  \cite{fannes1992finitely,SCHOLLWOCK1996, SCHOLLWOCKDMRG2011}. Usually in gapped one-dimensional quantum systems, the correlators take the Ornstein-Zernike form $\langle S_{m'}^z S_{m}^z \rangle \sim \exp{(-|m-m'|/\xi)}/\sqrt{|m-{m'}|}$ \cite{SCHOLLWOCKDMRG2011}. We fit to this form in order to extract correlation lengths for the hybrid chain. Figure~\ref{fig:fig3}(c) shows the correlation length as a function of $J_2/J_3$. We find the correlation length decreases quasiasymptotically toward the AKLT value (denoted by the green dotted line), then crosses it and settles below. The inset of Fig.~\ref{fig:fig3}(c) shows the correlation length of the BLBQ Hamiltonian [Eq.~\ref{eq:BLBQ}] as a function of $\beta$, with a clear minimum at the AKLT point 
$\beta=-\tfrac{1}{3}$ \cite{SCHOLLWOCK1996}. By matching the correlation length of our hybrid spin chain to that of the BLBQ model, we extract an effective $\beta_{\mathrm{eff}}$, shown in Fig.~\ref{fig:fig3}(d). While the correlation length as a function of $\beta$ for BLBQ model [inset in Fig.~\ref{fig:fig3}(c)] is not a single-valued function, we choose the branch $|\beta|<1/3$ because for  $J_2/J_3 \ll 1$ we approach the Heisenberg model from perturbation theory analysis for a hybrid spin dimer. The comparison is reliable only for $J_2/J_3 \lesssim 0.9$ [Fig.~\ref{fig:fig3}(c) data above the green dotted line], where the correlation length remains larger than $\xi \ge  \ln(3)^{-1}$ for the mapping to be well defined. By tuning the ratio $J_2/J_3$, we can effectively drive the system towards the AKLT point, albeit with a continuous and subtle deviation from the exact AKLT state. The correlation length dropping below the AKLT value of $ \xi \approx 0.91$ serves as an indicator of this deviation. We further compute the string order parameter (SOP) given as\cite{stringorder_Nijs,Satoshi_SOP_1992}
\begin{equation}
    \mathcal{O}^z =  - \lim_{|m-m'|\rightarrow \infty} \langle S^z_m \exp{(i\pi \sum_{m < n < m'} S^z_n)} S^z_{m'} \rangle,
\end{equation}
where we only consider spin operators on spin-1 sites. The SOP as well as the excitation gap are depicted in Fig.~\ref{fig:fig4} as a function of $J_2/J_3$ calculated for a chain of length $L=20$ using periodic boundary conditions. Indeed, the SOP remains finite and very close to the AKLT value of $\frac{4}{9}$ for almost all parameters, decreasing slightly only towards the Heisenberg point at $J_2/J_3\to 0$.

\subsection{Realization using nanographene flakes}

\noindent Using atomistic modeling, we now discuss how hybrid spin-chain models can be realized using nanographenes. Figure~\ref{fig:fig1}(a) presents a nanographene-based hybrid spin-1/spin-$\frac 12$ model consisting of two types of triangular graphene quantum dots with zigzag edges. The larger (red) triangles consist of $N=22$ carbon atoms and are known as triangulenes\cite{pavlivcek2017synthesis,mishra2019synthesis}. The smaller (blue) triangles contain $N=13$ carbon atoms and are known as phenalenyls\cite{MandalPhenalenyl2022,kubo2015phenalenyl}. Both triangulene and phenalenyl are open-shell nanographenes. Triangulene hosts two unpaired electrons that form a triplet ground state with $S=1$ due to the exchange interaction $J_1$, which is proportional to the onsite Coulomb interaction~\cite{Ezawa2008}. In contrast, phenalenyl hosts a single unpaired electron, yielding a ground state with $S=\frac{1}{2}$. The unpaired electrons of both structures localize at the edges \cite{Rossier2007, GuccluPRL2009, Potasz2010ZeroEnergy,potasz2012}. 
When triangles are connected in a manner conforming to the underlying triangular lattice, as depicted in Fig.~\ref{fig:fig1}(a), they couple antiferromagnetically via the superexchange mechanism\cite{Saleem2024}. From fermionic model calculations (see Appendix~\ref{app:Atomistic Model}–\ref{app:RealizationUsingNano}) combined with perturbation theory for a single spin-1 dimer, i.e., a spin-1-$\frac 12$-$\frac 12$-1 configuration as depicted in the inset of Fig.~\ref{fig:fig2}(b), we are able to determine the parameters for effective spin-$\frac 12$ model: $J_2 = 31.2$~meV and $J_3 = 58.7$~meV. This gives $J_2/J_3\approx0.5$. Moreover, a ferromagnetic exchange coupling between two electrons on the large triangles of $J_1 = -459$ meV ensures that triangulene can be treated as a spin-1 object as $|J_1| \gg J_2, J_3$ (see Appendixes~\ref{app: SpinParams} and \ref{app:RealizationUsingNano} for details). Further, we compute the many-body spectrum as a function of the number of spacers in a dimer. We find for $N_s=0,2,4,6$ the corresponding values for the effective BLBQ model  $J_{\mathrm{eff}} = 17.83, 10.98, 8.56, 6.84$ meV and $\beta_{\mathrm{eff}} = -0.067, -0.207, -0.232, -0.247$. Thus, increasing the number of spacers does reduce the effective coupling (and thus the Haldane gap) between the spin-1 sites, but also increases $\beta_{\mathrm{eff}}$, which leads to a reduction in the correlation length. The energy scale from a dimer without spacers $N_s=0$ to $N_s=6$ reduces almost three times, while the effective biquadratic term increases (in absolute value) almost four times. 

\section{Discussion}

\noindent We have demonstrated that inserting pairs of spin-$\frac{1}{2}$ spacers between spin-1 sites provides a powerful and general microscopic route to engineer and continuously tune unconventional biquadratic interactions in spin-1 chains. A perturbative analysis shows that the spacers lock into singlets for $J_2 \ll J_3$ and, via fourth order, generate an effective bilinear-biquadratic coupling. Using DMRG simulations, we uncover clear signatures of Haldane physics across a broad range of coupling ratios $J_2/J_3$—a finite excitation gap, nonzero string order, and spin-$\frac{1}{2}$ edge quasiparticles—with a correlation length that can be tuned smoothly toward the AKLT value. When $J_2 \ll J_3$, the hybrid spin model maps onto the BLBQ model, $H^{\rm eff}_{\rm BLBQ}= J_{\rm eff}\sum\limits_{i=1}^L\left[\Vec{S}_i\cdot\Vec{S}_{i+1}-\beta_{\rm eff}\left(\Vec{S}_i\cdot\Vec{S}_{i+1}\right)^2\right]$, so each bulk piece corresponds to a single effective spin-1 site. 

Finally, by combining atomistic calculations with our spin model, we propose a realistic solid-state platform for realizing this physics based on nanographene chains composed of 22-atom (spin-1) and 13-atom (spin-$\frac{1}{2}$) flakes. The calculated exchange ratio $J_2/J_3 \approx 0.5$ places these structures precisely in the optimal location—close to the AKLT form  — thereby providing a practical and tunable blueprint for engineering quantum spin liquids and topological states in bottom-up synthesized nanographene architectures.

Beyond the $N_s=2$ hybrid spin-1/spin-$\frac 12$ chain, our calculations reveal that increasing the number of spin-$\frac{1}{2}$ spacers, $N_s$, provides an additional tuning knob to engineer the effective BLBQ coupling. As shown in Appendix~\ref{app:DifferentNs}, adding more spacer pairs systematically enhances the magnitude of $\beta_{\mathrm{eff}}$ (while simultaneously reducing $J_{\mathrm{eff}}$), thereby driving the system closer to the AKLT form and shortening the correlation length of edge excitations. 

Recent advances in the on-surface synthesis of nanographene spin chains~\cite{FaselNaturespinchain,zhao2024tunable,Fu2025,yuan,Su2025spinonehalf,Sun2025OnSurfacespinonehalf} provide a promising platform to realize the hybrid spin chain proposed here. In particular, triangulene-based spin-1 units and phenalenyl-derived spin-$\frac{1}{2}$ units can, in principle, be incorporated into the same molecular framework, offering a chemically tunable route to engineer exchange pathways. The exchange ratios $J_2/J_3 \sim 0.5$ obtained from our atomistic simulations fall within a regime that appears experimentally accessible. Moreover, variations in edge passivation, linker chemistry, or flake size could allow systematic tuning of the effective couplings and exploration of the broader BLBQ phase diagram in nanographene-based spin arrays.

Finally, our approach to engineering biquadratic interactions via spin-1/2 spacers offers a flexible framework that can extend beyond one-dimensional spin-1 chains, promising new opportunities for designing quantum phases in higher-dimensional systems, such as triangular or kagome lattices, where tunable $\beta_{\mathrm{eff}}$ could stabilize exotic states like spin nematics or chiral spin liquids, building on recent theoretical proposals for 2D topological phases~\cite{BLBQInteractions2D2023Pohle,BLBQinteractions2DHoneycomb2024Mashiko,PRRPohle2024ChiralSpinLiquid}. We also noticed a recent theoretical proposal for realization of spin-1 chains using a chain of bilayer graphene quantum dots \cite{cns7-ynzy}.

\begin{acknowledgments}
Y.S. thanks Katarzyna Sadecka for valuable discussions and support. Y.S. and M.C. acknowledge funding by the Return Program of the State of North Rhine-Westphalia. W.P. and P.P. thank M. Kupczy\'nski for supporting calculations and constructive discussion. Calculations were partially performed using the Wroc\l aw Center for Networking and Supercomputing WCSS grant No. 317. Y.S. acknowledges financial support by the Deutsche Forschungsgemeinschaft (DFG, German Research Foundation) through the Würzburg-Dresden Cluster of Excellence ctd.qmat – Complexity, Topology and Dynamics in Quantum Matter (EXC 2147, project-id 390858490).
\end{acknowledgments}

\appendix

\section{Perturbation Theory} 
\label{appsub:MethodsPT}
Here we outline the perturbation theory analysis used for the dimer in the main text to analyze the limit of $J_2/J_3\ll1$. In this limit, the unperturbed Hamiltonian $H_0=H_{J_3}$, and  the perturbation $H'=H_{J_2}$. We specifically employ Schrieffer-Wolff perturbation, which separates the Hilbert space into low-energy $m$ and high-energy $l$ basis states. The low-energy states are chosen to be the ones in which the spin-$\frac 12$ spacers form a singlet state while the spin-1 part of the many-body state is unrestricted. Thus, the high-energy basis states $l$ are the ones with the spacers in triplet states. The zero-order energy gap between these states is $\Delta E=J_3$. There is no contribution to first order, that is $H'_{m,m'}=0$. This allows us to write the second-, third-, and fourth-order corrections as
\begin{subequations} 
\begin{align}
&H^{(2)}_{m,m'}=
-\frac{1}{J_3} \langle m| (H')^2 |m'\rangle ,
\label{eq:H2} \\
&H^{(3)}_{m,m'}=
\frac{1}{J_3^2} \langle m| (H')^3 |m'\rangle ,
\label{eq:H3} \\
&H^{(4)}_{m,m'}=
\frac{2}{J_3} \sum_{m''}H^{(2)}_{m,m''}H^{(2)}_{m'',m'} 
-\frac{1}{J_3^3} \langle m| (H')^4 |m'\rangle.
\label{eq:H4}
\end{align}
\end{subequations}
Thus, we need to evaluate averages of powers of $H'$ for the singlet spacer states. Using the relation
$s^is^j=\frac 14 \delta_{ij}+\frac{i}{2}\epsilon_{ijk}s^k$ for spin-1/2 operators
and $\epsilon_{ijk}S^iS^j=[S^{k+1},S^{k+2}]=iS^k$ for the spin-1's we get,
\begin{align}
\label{eq:recursion}
(H'/J_2)^2=
1-\frac 12 (H'/J_2) +2\sum_{ij}S^i_1S^j_2s^i_1s^j_2.
\end{align}Taking the average for spacer singlet states, one obtains
\begin{align}
H^{(2)}
=\frac{J_2^2}{2J_3} \mathbf{S}_1\cdot\mathbf{S}_2 - \frac{J_2^2}{J_3}.
\end{align}
Thus, to second-order correction, we get an antiferromagnetic 
coupling with constant $J^{(2)}=\frac{J_2^2}{2J_3}$ and a term proportional to the
identity, which has no direct impact. 
Multiplying Eq.~\ref{eq:recursion} by $H'/J_2$ yields results of similar structure as
the right-hand side of Eq.~\ref{eq:recursion} itself, thus giving rise to recursion relations 
that allow one to easily break down higher powers of $H'$. 
To third and fourth order, the Heisenberg coupling is corrected by $J^{(3)}=\frac{3J_2^3}{4J_3^2}$ and $J^{(4)}=-\frac{J_2^4}{8J_3^3}$, but the fourth order also generates a biquadratic term with $\beta_\text{eff} J_{\text{eff}}=\frac{J_2^4}{2J_3^3}$.

\section{DMRG calculations}
\label{appsub:DMRG}
The numerical analysis of spin Hamiltonians was performed using the finite density matrix renormalization
group (DMRG) method with single-site optimization implemented within the TenPy library \cite{tenpy}. The model is solved as a spin-1/2 chain, where spin-1 sites are realized by enforcing a large ferromagnetic coupling between two spin-1/2 sites. The ferromagnetic coupling is obtained from a fit of the fermionic model to the spin model. We get $J_1 = -459$ meV, $J_2 = 31.2$ meV, and $J_3 = 58.7$ meV. Clearly, $J_2/J_3 \sim 0.5$ and $|J_1| \gg J_2,J_3$. For calculations as a function of $J_2/J_3$, we fix $J_3 = 58.7$ meV and $J_1$ is increased, $J_1 = -459$ eV, while varying $J_2$.   We have set the maximal bond dimension to $\chi_{max} = 3000$ for the hybrid spin $\{ 1 \frac{1}{2} \}$ Hamiltonian, and $\chi_{max} = 2000$ for the BLBQ Hamiltonian, and performed at least 20 sweeps. The error bars on the numerical results are smaller than the data points.

\section{Full Atomistic Model} 
\label{app:Atomistic Model}
The Hubbard Hamiltonian for graphene nanostructures is given by
\begin{equation}
    H_{\mathrm{hub}}= \sum\limits_{i,l,\sigma}t_{il\sigma} c_{i\sigma}^\dag c_{l\sigma}
   +\frac{1}{2\kappa}U\sum\limits_{i,\sigma\neq\sigma'} c^\dag_{i,\sigma}c^\dag_{i,\sigma'}c_{i,\sigma'}c_{i,\sigma}, 
     \label{eq:Hub} 
\end{equation}
$c^\dagger_{i,\sigma}/c_{i,\sigma}$ creates/annihilates a $p_z$ electron on site $i$ (where sites correspond to the position of carbon atoms as depicted in Fig.~\ref{fig:fig1}(a)) with spin $\sigma$, $t_{il\sigma}$ are hopping elements, we take $t=-2.8$ eV as the nearest-neighbor (NN) hopping, $t'=-0.1$ eV as the next-nearest-neighbor (NNN) hopping and $t_3=-0.07$ eV is the next-next-nearest-neighbor (NNNN) hopping. We consider only the onsite (Hubbard) interaction $U=17.311$ eV screened by a dielectric constant $\kappa=3.5$ because these parameters capture the proper magnitudes of exchange couplings as well as the correct many-body spectrum for spin-1/2 and spin-1 chains\cite{Saleem2024,FaselNaturespinchain,yuan}. In these systems, genuine long-range Coulomb interactions are heavily screened due to the metallic substrate. As a reference for solving the interacting many-body problem, we first perform Hartree-Fock (HF) calculations. The HF equation is as\cite{GuccluPRL2009,potasz2012, saleemquantumsimulator,Saleem2024}
\begin{equation}
    H_{\mathrm{MF}}= \sum\limits_{i,l,\sigma}t_{il} c_{i\sigma}^\dag c_{l\sigma}
   +\frac{U}{\kappa}\sum\limits_{i,\sigma\neq\sigma'}
   \left(\rho_{ii\sigma'}-\frac{1}{2}\right)c^\dag_{i\sigma}c_{i\sigma},   
     \label{eq:HFHubbard}
    \end{equation}
with
\begin{equation}
       \rho_{ii\sigma'} =\sum_\lambda a^*_{\lambda i\sigma'}a_{\lambda i\sigma'}n_{\lambda\sigma'},
    \label{eq:DensityMatrix}
\end{equation}
where $n_{\lambda\sigma}$ are  occupations from the previous HF iteration step for state $\lambda$ with spin $\sigma$. HF calculations already provide a qualitatively correct spectrum of states near the Fermi level (in accordance with Lieb's theorem). To get better quantitative results, as well as obtain excited states we perform complete active space (CAS) calculations of the 12 (times 2 for spin degeneracy) HF states closest to the Fermi level. To this end, we diagonalize the many-body Hamiltonian rotated into the HF basis: 
\begin{equation}
\begin{split}
 H_{\mathrm{hub}}&= \sum\limits_{p,\sigma}\epsilon^{HF}_{p\sigma}b_{p\sigma}^\dagger b_{p\sigma}-\sum\limits_{p,q,\sigma} \tau_{pq\sigma}b_{p\sigma}^\dagger b_{q\sigma}\\
 &+ \frac{1}{2}\sum\limits_{p,q,r,s,\sigma,\sigma'}\braket{pq|V|rs}b_{p\sigma}^\dagger b_{q\sigma'}^\dagger b_{r\sigma'} b_{s\sigma},
    \end{split}
    \label{eq:hubHamHFbasis}
\end{equation}
where, $b^\dagger_{p,\sigma}/b_{p,\sigma}$ creates/annihilates a particle on Hartree-Fock (HF) state $p$ with spin $\sigma$, and $\epsilon^{HF}_{p\sigma}$ corresponds to the energy of this state. The double counting corrections $\tau_{pq\sigma}$ are given explicitly as
\begin{equation}
\tau_{pq\sigma} = \sum_{m\sigma'}\left(\braket{pm|V|mq}-\braket{pm|V|qm}\delta_{\sigma\sigma'}\right)n_{m\sigma'},
\label{eq:DC}
\end{equation}
where $n_{m\sigma'}$ measures occupation. The Coulomb matrix elements $\braket{pq|V|rs}$ are given by the rotation:
\begin{equation}
\braket{pq|V|rs} = \frac{U}{\kappa}\sum_{i,\sigma\neq\sigma'}a^*_{pi\sigma}a^*_{qi\sigma'}a_{ri\sigma'}a_{si\sigma},
\label{eq:CME}
\end{equation}
where, $a_{pi\sigma}$ are eigenvectors obtained by solving the HF equation. 

\section{From Fermionic to Spin Model} 
\label{appsub:FermionictoSpin}
As we have shown in our previous work \cite{Saleem2024},  three states below and three states above six quasi-degenerate energy states, have to be included in order to take into account the superexchange mechanism. Alternatively, this effect might be included by enhancing NNNN hopping value of $t_3=-0.33$ eV \cite{Henriques2023t3paper}.
 By doing so, we can confine ourselves to only states within the degenerate shell at the Fermi level and reduce the size of our Hilbert space dramatically. We then rewrite Eq.~\ref{eq:Hub} as 
\begin{equation}
    H_{t_3}= \sum_{\Delta}H_\Delta+\sum_{\Delta\neq\Delta'}V_{\Delta\Delta'},
    \end{equation}
where 
\begin{equation}
    H_\Delta= \sum\limits_{i,l\in S_{\Delta}}\sum_{\sigma}t_{il\sigma} c_{i\sigma}^\dag c_{l\sigma}
   +\frac{1}{2}U\sum\limits_{i\in S_{\Delta}} \sum_{\sigma\neq\sigma'}c^\dag_{i,\sigma}c^\dag_{i,\sigma'}c_{i,\sigma'}c_{i,\sigma}
    \end{equation}
    and
    \begin{equation}
    V_{\Delta\Delta'}= t_3\sum\limits_{i\in S_\Delta}\sum_{l\in S_{\Delta'} }\sum_{\sigma}c_{i\sigma}^\dag c_{l\sigma}+h.c.,
    \end{equation}
where $\Delta=1,2,3,4$ corresponds to the four triangles (big-small-small-big), and $S_{\Delta}$ corresponds to the set of sites (location of Carbon atoms) belonging to each triangle. That is, we decomposed the Hamiltonian into the Hamiltonian of individual triangles (effective spins) denoted by $H_\Delta$ and the coupling between these triangles (or effective spins) denoted by $ V_{\Delta\Delta'}$. The solutions of $H_\Delta$ are the solutions of individual triangular graphene quantum dots with zigzag edges (TGQD)s. It is well known that these TGQDs contain degenerate states at the Fermi level, a phenomenon resulting from sublattice imbalance. The number of degenerate states is equal to the number of atoms along the edge minus 1 (i.e. $N_{edge}-1$) \cite{Rossier2007,Potasz2010ZeroEnergy}. Thus for the larger triangle (triangulene) there are two degenerate shell states at the Fermi level, while there is only one for the smaller triangle (Phenalenyl). Furthermore, the wavefunctions for these states are C3 symmetric and localized at the edges of the TGQD\cite{Rossier2007, Henriques2023t3paper, saleemquantumsimulator, Saleem2024, potasz2012, Potasz2010ZeroEnergy}. Thus, by confining ourselves to these degenerate shell states, we can then rotate our Hamiltonian in Eq.~\ref{eq:Hub} to the basis of these C3 symmetric edge modes to get 
\begin{equation}
\begin{aligned}
H_{\mathrm{C3}}&= \sum\limits_{\Delta,\Delta',\alpha,\alpha',\sigma}\tau_{\Delta\alpha,\Delta'\alpha'} d^\dagger_{\Delta,\alpha,\sigma}d_{\Delta',\alpha',\sigma} \\
   &+\frac{1}{2}\sum\limits_{\Delta,\alpha_1,\alpha_2,\alpha_3,\alpha_4,\sigma,\sigma\neq\sigma'}U_{\Delta} d^\dagger_{\Delta,\alpha_1,\sigma}d^\dagger_{\Delta,\alpha_2,\sigma'}d_{\Delta,\alpha_3,\sigma'}d_{\Delta,\alpha_4,\sigma}, 
\end{aligned}
\label{eqn:C3_Ham}
\end{equation}
where $d^\dagger_{\Delta,\alpha,\sigma}/d_{\Delta,\alpha,\sigma}$ creates or annihilates a particle on triangle $\Delta$ and mode $\alpha$ with spin $\sigma$. For triangulene which we will denote with $B$ (the big triangle) there are two modes which we will denote with $\alpha=\pm$, while for Phenalenyl which we will denote with $S$ (the small triangle) there is a single mode which we will drop the index $\alpha$. Due to the local nature of the onsite Hubbard $U$ and the C3 symmetry of the edge modes we find that the Coulomb elements $U_\Delta$ depend only on $\Delta$ and satisfy the angular momentum conserving selection rule $\alpha_1+\alpha_2=\alpha_3+\alpha_4$. That is, direct, exchange and onsite interactions between electrons on triangulene are all equal. We compute this to be $U_B=0.459$ eV and $U_S=0.8243$ eV (already screened appropriately). The hopping between the edge modes are $\tau_{\Delta\alpha,\Delta'\alpha'}$ with the non-zero elements being $\tau_{B\pm,S}=-0.0712 \pm0.0411i$ eV  and $\tau_{S,S}=0.11$ eV. 

In this basis, the basis states correspond to effective spin-1/2 states of each individual triangles. Thus we can now interpret $J_1=-U_B$ as the ferromagnetic exchange coupling between the two degenerate shell modes of triangulene \cite{Ezawa2008}. While $J_2,J_3$ correspond to exchange coupling between the triangulene-Phenalenyl and Phenalenyl-Phenalenyl respectively. To obtain $J_2$ and $J_3$ we use second-order perturbation theory and determine $J_2 =\frac{2|\tau_{B\pm,S}|^2}{U_S}+\frac{2|\tau_{B\pm,S}|^2}{U_B}= 31.2$ meV, $J_3 =\frac{4|\tau_{S,S}|^2}{U_S}=  58.7$ meV. We neglect a tiny energy difference between edge modes on triangulene and Phenalenyl, which does not affect the results significantly. A comparison between full atomistic model, a simplified $C_3$ Fermionic model and hybrid with $J_2$ and $J_3$ parameters for the hybrid spin-1/-$\frac 12$ dimer, a system consisting of two triangulenes connected by two Phenalenyls, is shown in Fig.~\ref{fig:figS1}.
\section{Determination of spin parameters from atomistic model}
\label{app: SpinParams}
\noindent We outline how the spin parameters $J_2$, $J_3$ are determined for the nanographene spin chain. We focus on $J_3$; $J_2$ follows analogously. To extract $J_3$, we first consider a two site spin-1/2 Hamiltonian given by 
\begin{equation}
    H_{s=\frac{1}{2}} = J_3\textbf{s}_1\cdot\textbf{s}_2.
\end{equation}
In terms of raising and lowering operators, we can rewrite this as
\begin{equation}
    H_{s=\frac{1}{2}} = J_3\left[\frac{1}{2}\left(s^+_1s^-_2+s^-_1s_2^+\right)+s_1^zs_2^z\right].
\end{equation}
Each spin-1/2 site has two possible states, $\ket{\uparrow}$ and $\ket{\downarrow}$. For the two sites, there are a total of four states, which reduce to two when restricting to the $(s_1^z+s_2^z)=0$ subspace. This yields a 2x2 Hamiltonian
\begin{equation}
     \textbf{H}_{s=\frac{1}{2}} = \frac{J_3}{4} \begin{pmatrix}
-1 & 2 \\
2 & -1
\end{pmatrix}.
\label{eq:2sitespin12}
\end{equation}
$J_3$ corresponds to an effective exchange coupling between two phenalenyls. Now using Eq.~\ref{eqn:C3_Ham} and restricting ourselves to one degenerate state of each phenalenyl, we can solve this problem analytically using configuration interaction. That is, there is one unpaired electron for each phenalenyl. For $S_z=0$ there are four configurations in total: $\ket{1}=c^{\dag}_{1\uparrow}c^{\dag}_{2\downarrow}\ket{0}$, $\ket{2}=c^{\dag}_{1\downarrow}c^{\dag}_{2\uparrow}\ket{0}$, $\ket{3}=c^{\dag}_{1\downarrow}c^{\dag}_{1\uparrow}\ket{0}$, $\ket{4}=c^{\dag}_{2\downarrow}c^{\dag}_{2\uparrow}\ket{0}$. The CI Hamiltonian in this basis is given as
\begin{equation}
    H^{J_3}_{\mathrm{CI}} = \begin{pmatrix}
0 & 0 & -\tau_{SS} & \tau_{SS} \\
0 & 0 & \tau_{SS} & \tau_{SS} \\
-\tau_{SS} & \tau_{SS} & U_S & 0 \\
\tau_{SS} & \tau & 0 & U_S
\end{pmatrix}
,
\end{equation}
where $\tau_{SS}$ and $U_S$ are defined in Appendix.~\ref{appsub:FermionictoSpin}. They correspond to hopping between the zero-energy modes of the phenalenyls ($\tau_{SS}$) and the energy cost of putting two electrons on the same edge mode of a single phenalenyl ($U_S$). The two lowest energy states correspond to the singly occupied configurations $\ket{1}$, $\ket{2}$. We now use perturbation theory by treating the singly occupied configurations as our unperturbed states. The C3 Hamiltonian is first rewritten in the perturbative form as
$H_{C3}=H_0+\lambda V$ where,
\begin{equation}
    V= \sum\limits_{\Delta,\Delta',\alpha,\alpha',\sigma}\tau_{\Delta\alpha,\Delta'\alpha'} d^\dagger_{\Delta,\alpha,\sigma}d_{\Delta',\alpha',\sigma}.
\end{equation}
Second-order perturbative corrections are given by
\begin{equation}
    H_{nk}^{(2)} = -\sum_m \frac{\langle n | V | m \rangle \langle m | V | k \rangle}{\Delta E_m^{(1)}},
    \label{eq:SW2ndOrder}
\end{equation}
where $n,k$ correspond to the unperturbed states, and $m$ corresponds to the rest of the states (in this case configurations 3, 4), and $\Delta E_m=U_S$. We note that first-order perturbation theory gives no contribution. Second-order perturbation theory gives the matrix
\begin{equation}
    H^{(2)}_{J_3} = \begin{pmatrix}
-\frac{|\tau_{SS}|^2}{U_S} & 2\frac{|\tau_{SS}|^2}{U_S}\\
2\frac{|\tau_{SS}|^2}{U_S}& -\frac{|\tau_{SS}|^2}{U_S}\\
\end{pmatrix}
.
\end{equation}
Direct comparisons between this equation and Eq.~\ref{eq:2sitespin12} gives us
\begin{equation}
    J_3=4\frac{|\tau_{SS}|^2}{U_S}. 
\end{equation}
Following an analogous procedure, one can derive $J_2$ giving the result here as
\begin{equation}
    J_2 =\frac{2|\tau_{B\pm,S}|^2}{U_S}+\frac{2|\tau_{B\pm,S}|^2}{U_B}.
\end{equation}

\section{Realization of hybrid spin-1/-$\frac 12$ chain using nanographenes}
\label{app:RealizationUsingNano}
\begin{figure}
\includegraphics[width=\linewidth]{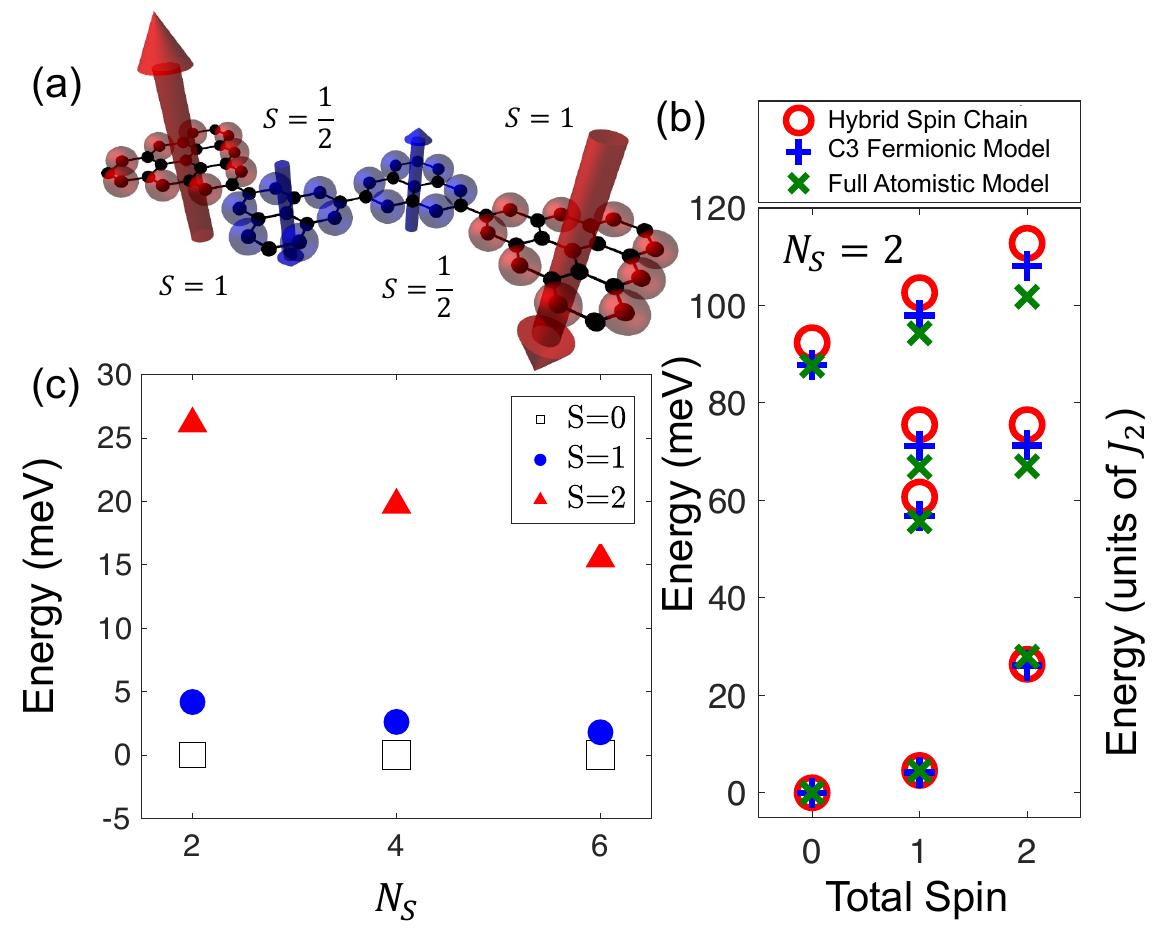}
    \caption{(a) Hybrid spin-1/-$\frac 12$ dimer made of the atomistic alternating nanographene spin-1's and two spin 1/2's, $N_S=2$. The balls represent carbon atoms, with the line connecting carbon atoms representing nearest-neighbor hopping. We highlight the localization of the effective spin of these nanographenes along the edge and show that each nanographene can be represented by an effective spin. In (b), we show the low-energy many-body spectrum of a dimer labelled by total spin $S$ for different theoretical models with $N_S= 2$. (c) Low-energy many-body spectrum from the C3 Fermionic model for alternating nanographene spin dimers with increasing spin-1/2 buffers $N_s$.
} 
\label{fig:figS1}
 \end{figure}

\noindent As we discussed in the main text, hybrid spin-chain models can be experimentally realized using nanographenes. Fig.~\ref{fig:figS1}(a) presents a dimer of the hybrid spin-1/-$\frac 12$ model consisting of four connected triangular graphene quantum dots with zigzag edges. The edge localization of the unpaired electrons is shown using colored circles on top of each edge site.  
We analyze three models for the hybrid spin dimer, described in Appendix~\ref{app:Atomistic Model} and Appendix~\ref{appsub:FermionictoSpin}. The first involves performing full atomistic calculations within a Hubbard model, increasing the size of the Hilbert space CAS until the low-energy states converge. The second model, known as the C3 Fermionic model, is an approximate model, whereby manually increasing the value of the next-next-nearest-neighbor hopping $t_3$ that connects the edge modes of adjacent nanographene allows us to confine ourselves to only the degenerate shell near the Fermi level, reducing the size of the many-body Hilbert space \cite{Henriques2023t3paper,Saleem2024}. Using the C3 Fermionic model combined with perturbation theory, we are able to determine the three parameters: $J_1 = -459$ meV, $J_2 = 31.2$ meV, and $J_3 = 58.7$ meV. This gives $J_2/J_3\approx0.5$. $J_1$ is the ferromagnetic exchange coupling between the two electrons on the large triangles. We also have $|J_1| \gg J_2, J_3$, which ensures that triangulene can be treated as a spin-1 object, allowing us to directly map the Fermionic Hamiltonian into the hybrid spin-1/-$\frac{1}{2}$ model. Consequently, $J_1$ can be neglected in the hybrid spin model. We can then use $J_2, J_3$ parameters to diagonalize the Hamiltonian of the hybrid spin-1/-$\frac 12$ model defined in Eq.~(2) and Eq.~(3) of the main text for $N_s=2$. Fig.~\ref{fig:figS1}(b) shows the low-energy many-body spectrum for these three models. The same exact spin-ordering of many-body states is obtained. In particular, the three lowest states are close in energy—a singlet (the ground state) and a triplet (the first excited state)—while a quintuplet lies clearly higher in energy. This is the characteristic energy spectrum of a $S=1$ dimer. 

We further compute the many-body spectrum as a function of the number of spacers $N_S$. Fig.~\ref{fig:figS1}(c) shows the first three many-body states composed of a singlet, triplet, and quintuplet for $N_s=2,4,6$ obtained for the C3 Fermionic model. To gain insight into the calculations, we fit a 2-site spin-1 BLBQ model defined by Eq.~(1) to the low-energy manifold of states shown in Fig.~\ref{fig:figS1}(c). We find for $N_s=0,2,4,6$ the corresponding values of  $J_{\mathrm{eff}} = 17.83, 10.98, 8.56,6.84$ meV and corresponding values of $\beta_{\mathrm{eff}} = -0.067, -0.207, -0.232, -0.247$. Thus, increasing the number of spacers does reduce the effective coupling (and thus the Haldane gap) between the spin-1 sites, but also increases $\beta_{\mathrm{eff}}$. 

\begin{figure} [h]
    \centering
    \includegraphics[width=1\linewidth]{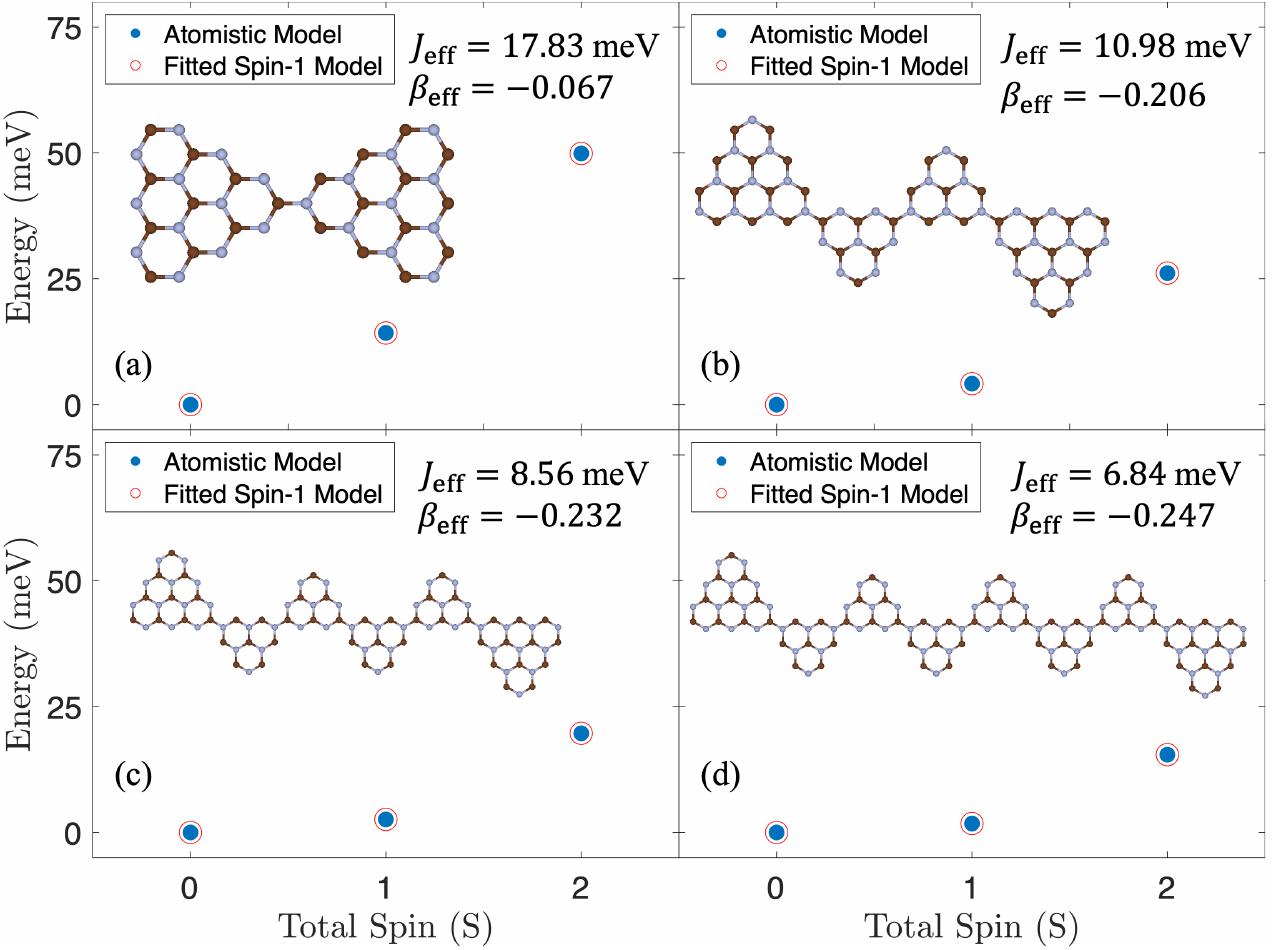}
    \caption{Dimers of the hybrid spin-1/-$\frac 12$ models with different numbers of spacers $N_s$ made of triangular graphene quantum dots with zigzag edges. (a) $N_s = 0$, (b) $N_s = 2$, (c) $N_s = 4$, (d) $N_s = 6$. $N_s$ controls an energy scale and biquadratic term $\beta_{\mathrm{eff}}$.}
    \label{fig:dimers_Ns}
\end{figure}
A comparison between the energy spectra obtained from the atomistic model and the fitted spin-1 model for dimers with different numbers of spacers is shown in Fig.~\ref{fig:dimers_Ns}. Parameters for effective spin-1 dimers are determined. The energy scale from a dimer without spacers $N_s=0$ Fig.~\ref{fig:dimers_Ns}(a) to $N_s=6$ Fig.~\ref{fig:dimers_Ns}(d) reduces almost three times, while the effective biquadratic term increases (in absolute values) almost four times.

\section{Convergence analysis of DMRG} 

\begin{figure} [h]
    \centering
    \includegraphics[width=0.9\linewidth]{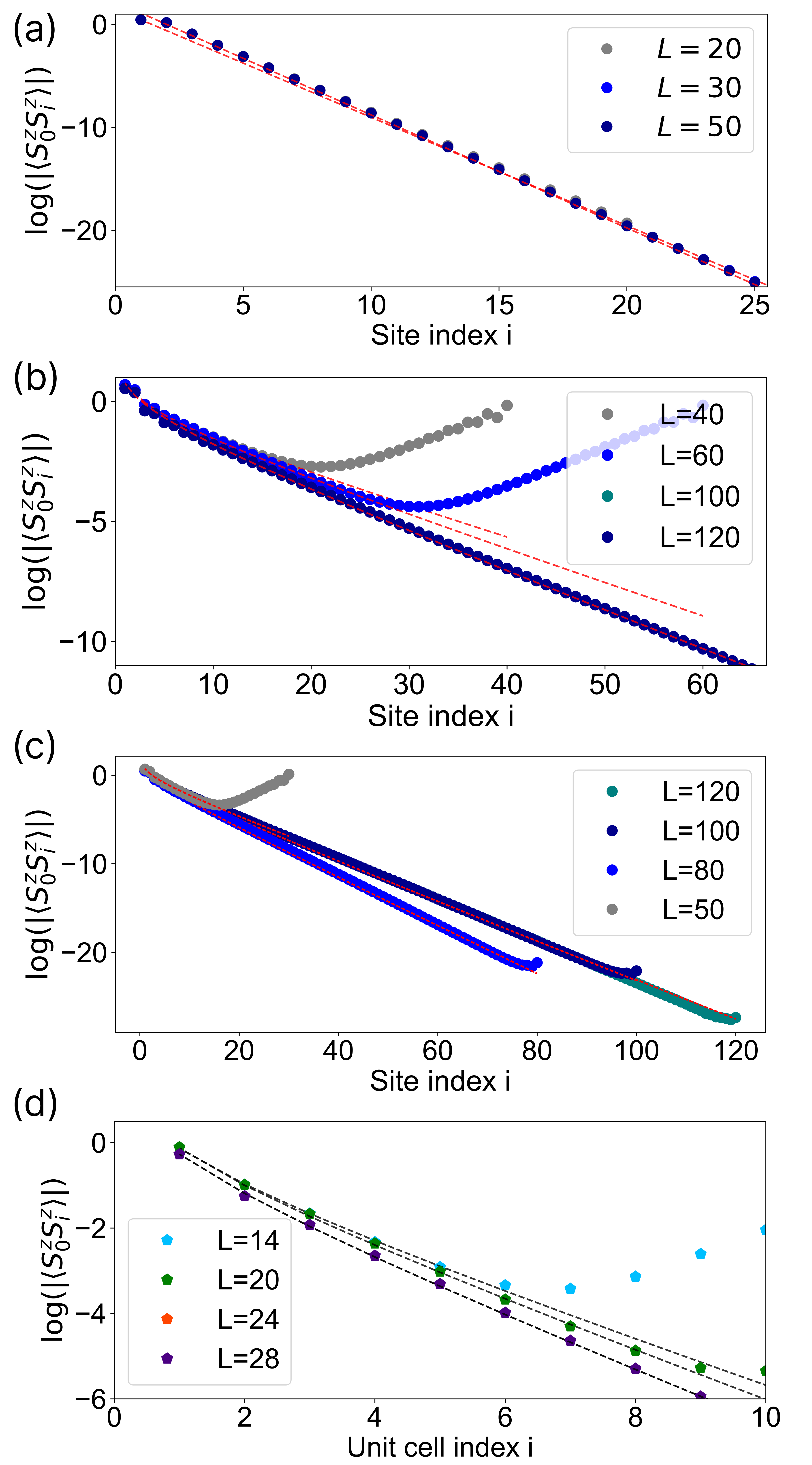}
    \caption{The convergence of spin-spin correlations of the ground state for four models: BLBQ models (a)-(c) and hybrid spin-1/-$\frac 12$ chain model (d). (a) The AKLT state,  $\beta_{\mathrm{eff}} =-\frac{1}{3}$, (b)  spin-1 Heisenberg model, $\beta_{\mathrm{eff}} =0$, (c) $\beta_{\mathrm{eff}} = 0.09$. (d) Hybrid spin-1/-$\frac 12$ chain model for $J_2/J_3 = 0.5$.}
    \label{fig:conv_CL}
\end{figure}
\noindent The convergence analysis presented here consists of two parts. The first concerns obtaining the correct ground state that minimizes the expectation value of the Hamiltonian, the second part involves obtaining accurate excited states. The primary parameter governing how well the Matrix Product State  (MPS) approximates the target state is the maximum bond dimension $\chi_{BD}$, which defines the size of the bond index connecting neighboring tensors in the chain. The spin chain models studied in this work do not require large bond dimensions, and ground state energy convergence is achieved for $\chi_{BD}<1000$. However, due to the chain lengths considered and the need to compute excited states, we set $\chi_{BD} = 2000$ for the BLBQ Hamiltonian and $\chi=3000$ for the hybrid spin-1/-$\frac 12$ Hamiltonian.

We also performed a convergence analysis with respect to chain length for the correlation length. These results are summarized in Figure \ref{fig:conv_CL}: for the hybrid spin model, convergence is reached at 24 unit cells, while for the BLBQ Hamiltonian with $\beta = 0$ and $\beta=0.09$, convergence is observed for $L=100$ sites. The correlation length of the AKLT state does not depend on the chain length.

\section{Spin densities of the hybrid spin-1/spin-$\frac 12$ chains from DMRG calculations} 

\begin{figure}
    \centering
    \includegraphics[width=\linewidth]{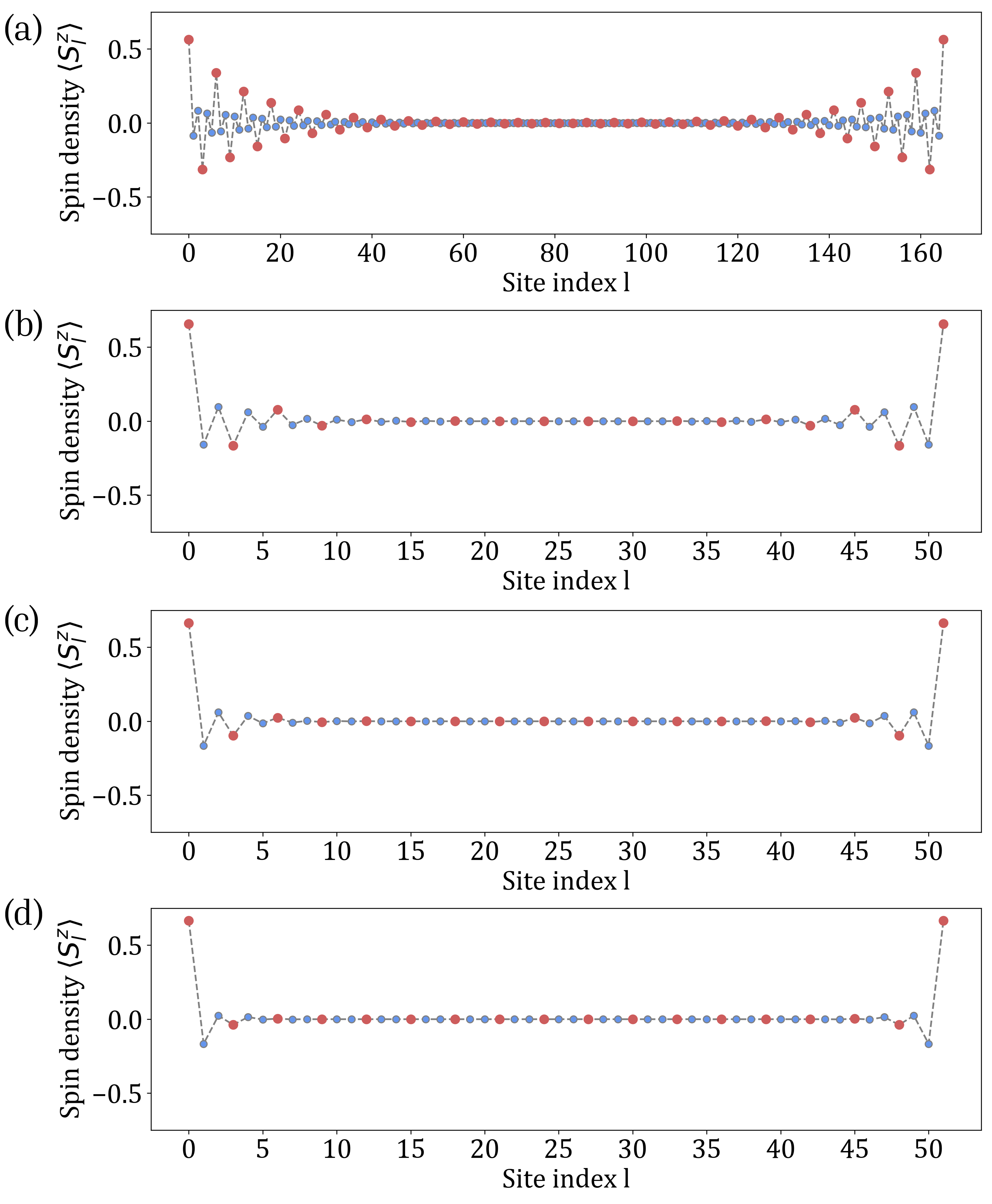}
    \caption{The spin densities of the ground state for the hybrid spin-1/-$\frac 12$ chain model with parameters (a) $J_2/J_3 = 0.2$, (b) $J_2/J_3 = 0.7$, (c) $J_2/J_3 = 1.2$, (d) $J_2/J_3 = 3.0$. An index $l=3(m-1)+i$, where $m$ is a unit cell index and $i=0,1,2$, with $i=0$ for a spin-1 and $i=1,2$ for spin-$\frac 12$'s.}
    \label{fig:spin-densities}
\end{figure}

\noindent In Fig.~\ref{fig:spin-densities} we show spin densities for the hybrid spin-1/-$\frac 12$ chain model with parameters (a) $J_2/J_3 = 0.2$, (b) $J_2/J_3 = 0.7$, (c) $J_2/J_3 = 1.2$, (d) $J_2/J_3 = 3.0$. Chain lengths are chosen to be $L = 55, 18, 18, 18$ respectively. The site index is $l=3(m-1)+i$ , where $m$ is a unit cell index and $i=0,1,2$, with $i=0$ for a spin-1 and $i=1,2$ for spin-$\frac 12$'s. The increase in localization at the first and last lattice sites and the reduction of edge density variations with the increase in $J_2/J_3$ is visible.

\section{The hybrid spin-1/spin-$\frac 12$ chain model with a different number of spacers $N_s$}
\label{app:DifferentNs} 
\begin{figure}
    \centering
    \includegraphics[width=1\linewidth]{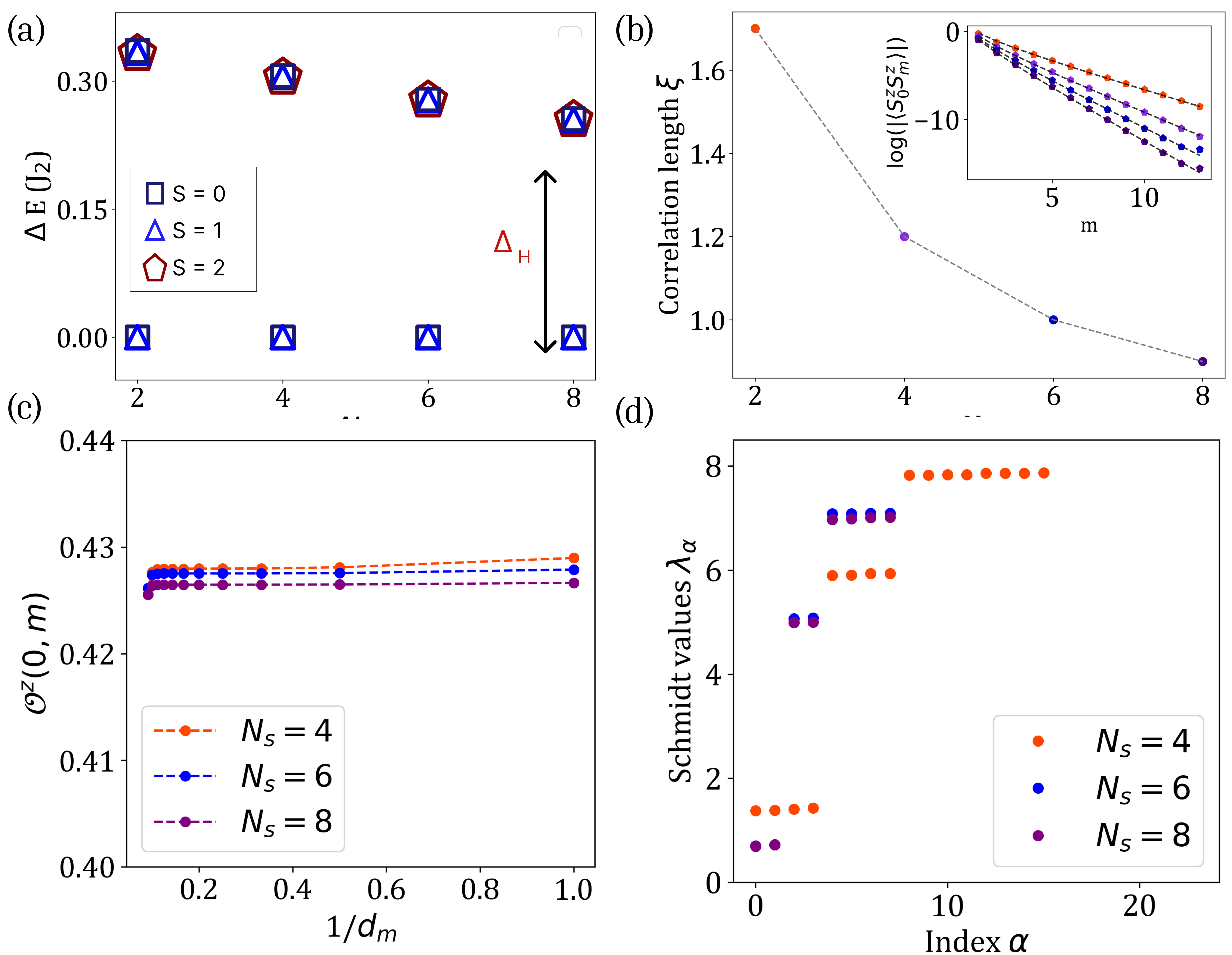}
    \caption{(a) The low energy spectrum of the hybrid spin-1/-$\frac 12$ chain models with different number of spacers $N_S = 2, 4, 6, 8$, black arrow denotes the Haldane gap $\Delta_H$. (b) Relation between the correlation length $\xi$ and number of spacers $N_S$, the inset shows spin-spin correlations fitted to the analytical formula $\ln{\left(\exp{(-|j-i|/\xi)}/\sqrt{|j-i|}\right)}$. (c) String order parameters and (d) entanglement spectra for three chains with the number of spacers $N_S = 4, 6, 8$. All results obtained for the parameter ratio $J_2/J_3 = 0.5$.}
    \label{fig:NS_energy_SOP_ES}
\end{figure}

We investigate the hybrid spin-1/-$\frac 12$ chain model with a different number of spacers $N_s$ using DMRG. Lengths of chains are chosen such that the results are converged, which correspond to $L=24$ for $N_S = 2$, and $L=16$ for $N_s=4,6,8$.
Fig.~ \ref{fig:NS_energy_SOP_ES} shows in (a) the low energy spectrum with a black arrow indicating the Haldane gap $\Delta_H$, (b) the correlation length $\xi$ obtained from fitting to the analytical formula $\ln{\left(\exp{(-|j-i|/\xi)}/\sqrt{|j-i|}\right)}$, (c) String order parameters and (d) entanglement spectra. All results obtained for $J_2/J_3 = 0.5$. We notice a slight decrease in the Haldane gap with an increase of the number of spacers $N_s$, while a drastic decrease in the correlation length. All the structures exhibit hidden antiferromagnetic order and double degeneracy of eigenvalues in their entanglement spectra for the cut between spin-1 and spin-$\frac 12$. 

\clearpage
\newpage



\bibliography{Main}

\clearpage
\newpage

\end{document}